# Shadow Mask Molecular Beam Epitaxy for In-Plane Gradient Permittivity Materials

S. Mukherjee[1], S. R. Sitaram[1], X. Wang[1], and S. Law[2,3,4,5,*]

[1]Department of Materials Science and Engineering, University of Delaware, 201 Dupont Hall, 127 The Green, Newark, DE 19716, USA

[2]Department of Materials Science and Engineering, The Pennsylvania State University, University Park, PA 16802, USA

[3]Materials Research Institute, The Pennsylvania State University, University Park, PA 16802, USA

[4]2D Crystal Consortium Materials Innovation Platform, The Pennsylvania State University, University Park, PA 16802, USA

[5]Penn State Institute of Energy and the Environment, The Pennsylvania State University, University Park, PA 16802, USA

[*]sal6149@psu.edu

**Abstract**

Infrared spectroscopy currently requires the use of bulky, expensive, and/or fragile spectrometers. For gas sensing, environmental monitoring, or other applications in the field, an inexpensive, compact, robust on-chip spectrometer is needed. One way to achieve this goal is through gradient permittivity materials, in which the material permittivity changes as a function of position in the plane. In this paper, we demonstrate the synthesis of infrared gradient permittivity materials using shadow mask molecular beam epitaxy. The permittivity of our material changes as a function of position in the lateral direction, allowing us to confine varying wavelengths of infrared light at varying horizontal locations. We see an electric field enhancement corresponding to a wavenumber gradient of ~650 cm$^{-1}$ to 900 cm$^{-1}$ over an in-plane gradient width of ~13 μm on the flat mesa of our sample. In addition, we see a wavenumber gradient of ~900 cm$^{-1}$ to 1250 cm$^{-1}$ over an in-plane gradient width of ~13 μm on the slope of our sample. These two different wavenumber gradient regions develop on two opposite sides of our material. This demonstration of a scalable method of creating an in-plane gradient permittivity material could be leveraged for the creation of a variety of miniature infrared devices, such as an ultracompact spectrometer.

In this paper, we demonstrate synthesis of infrared gradient permittivity materials using shadow mask molecular beam epitaxy. Permittivity of our material changes as a function of position in the lateral direction, allowing to confine varying wavelengths of infrared light at varying horizontal locations. We see electric field enhancement corresponding to a wavenumber gradient of ~650 cm$^{-1}$ to 900 cm$^{-1}$ over an in-plane gradient width of ~13 µm on the flat mesa of our sample. In addition, we see a wavenumber gradient of ~900 cm$^{-1}$ to 1250 cm$^{-1}$ over an in-plane gradient width of ~13 µm on the slope of our sample. These two different wavenumber gradient regions develop on two opposite sides of our material. This demonstration of a scalable method of creating in-plane gradient permittivity material could be leveraged for the creation of a variety of miniature infrared devices, such as an ultracompact spectrometer.

## 1. Introduction

Shadow mask molecular beam epitaxy (SMMBE) is a form of selective area epitaxy (SAE) which was developed to enhance the flexibility of conventional MBE.[1] As the name suggests, SMMBE uses a mask either directly fabricated on the substrate or placed in contact with the substrate. During film deposition, epitaxial layers are grown[2] on the substrate through apertures in the mask. In this way, the film grows only in the desired areas, removing the need for post-growth etching and the concomitant damage to the film. The use of a mechanical shadow mask for selective MBE growth was first demonstrated by Cho and Reinhart for dielectric waveguide fabrication, with the waveguide width set by the mask dimension.[3,4] Since then, SMMBE has been used for the growth of a wide range of materials and heterostructures.[5–20] Several studies[1,3,5] reported a shadowing effect near the mask edges in which elemental fluxes vary as a function of position. This effect is caused by imperfect mask edges combined with a non-zero angle between the effusion cells and surface normal. This results in a gradient of film thickness and/or composition near the mask edges. The steepness and the width of the gradient can be controlled by varying the mask thickness and/or the angle of the mask edges. In this paper, we demonstrate the potential of the SMMBE technique to create in-plane gradient permittivity materials (GPMs) by taking advantage of the shadowing effect.

A GPM is a material in which the permittivity varies as a function of location. Our aim is to synthesize in-plane GPMs, in which the permittivity varies in the horizontal/in-plane direction rather than in the vertical growth direction. In an in-plane GPM, different wavelengths of light can be confined[21] at different in-plane locations on the chip. This structure could be used to build an ultracompact on-chip spectrometer. We are interested in working in the infrared (IR) regime, so our GPMs are comprised of heavily silicon-doped indium arsenide (Si:InAs), which is known to be a good IR plasmonic material.[22–26] The permittivity of doped semiconductors can be modeled with the Drude formalism as shown in **Equation 1**:[27]

$$\varepsilon_{Drude} = \varepsilon_\infty \left(1 - \frac{\omega_p^2}{\omega^2 + i\omega\Gamma}\right) \quad (1)$$

Here, $\varepsilon_\infty$ is the high-frequency permittivity of the InAs, $\omega_p$ is the plasma frequency of the doped InAs, $\omega$ is the frequency of incident light, and $\Gamma$ is the electron scattering rate.[27] The plasma frequency, $\omega_p$, is related to the carrier density via **Equation 2**:[28]

$$\omega_p = \sqrt{\frac{ne^2}{\varepsilon_\infty \varepsilon_0 m^*(n)}} \quad (2)$$

Here, $n$ is the 3D carrier density, $e$ is the carrier charge, $\varepsilon_\infty$ is the high-frequency permittivity of the InAs, $\varepsilon_0$ is the permittivity of free space, and $m^*(n)$ is the effective mass of the carriers[29] which depends on $n$.[28] InAs can be very heavily doped, with $n$ reaching ~1×10$^{20}$ cm$^{-3}$, which leads to plasma wavelengths as short as ~5 μm,[27] which is the shortest demonstrated plasma wavelength among the III-V semiconductors. Using MBE to tune the carrier density in Si:InAs enables the plasma wavelength to be tuned across the mid-IR.[22,27,28]

Unfortunately, traditional MBE growth can only be used to change the carrier density in the growth direction by changing the silicon and/or the indium fluxes. Creating a GPM using traditional MBE would therefore require the growth of a very thick layer of Si:InAs in which the doping density slowly changes as a function of depth. The sample would then need to be cleaved and the GPM would be fabricated on the cleaved edge. This is technologically challenging, and the resulting devices would be extremely small. Current methods of in-plane GPM fabrication include ion irradiation patterning,[30] RF magnetron sputtering,[31] and patterned spin-on dopants.[32] These techniques have a variety of downsides, such as film damage[33] and contamination.[34]

To create our GPMs, we use the SMMBE technique. Rather than trying to minimize the flux gradients near the edges of the mask, we aim to enhance and control them. By creating flux gradients of both indium and silicon near the edges of the mask, we can control the permittivity of Si:InAs in the in-plane direction of the film. Each location will thus have a different carrier density, leading to a different plasma frequency, $\omega_p$, and ultimately to a different permittivity, $\varepsilon_{Drude}$. In this paper, we demonstrate the successful synthesis of a Si:InAs in-plane GPM using SMMBE. We show that we can get in-plane permittivity gradients in our Si:InAs film: on the flat mesa on the side where the silicon flux is shadowed, and on the film slope on the opposite side where the indium flux is shadowed. Using scanning near-field optical microscopy, we see an electric field enhancement corresponding to a wavenumber gradient of ~650 cm$^{-1}$ to 900 cm$^{-1}$ over an in-plane gradient width of ~13 μm on the flat mesa, and we see a wavenumber gradient of ~900 cm$^{-1}$ to 1250 cm$^{-1}$ over an in-plane gradient width of ~13 μm on the slope. In contrast to the flat mesa, we observe that the in-plane permittivity gradient on the slope is steeper and wider and reaches a higher maximum wavenumber of ~1250 cm$^{-1}$. Either of the in-plane GPMs on the flat mesa or the slope can be used to build an ultracompact on-chip spectrometer.

## 2. Results and discussion

To synthesize our samples, we used removable and potentially reusable shadow masks made of silicon purchased from Norcada (NX10500).[35] **Figure 1(a)** shows cross-sectional view of the mask. The mask is 200 μm thick and 1 cm × 1 cm in size with an aperture at the center that is 0.5 cm × 0.5 cm at the top and 0.528 cm × 0.528 cm at the bottom with 54.7° sidewalls, shown schematically in Figure 1(a). Here the mask is shown in its upside-down orientation, the way it is inserted into the MBE chamber for sample growth.

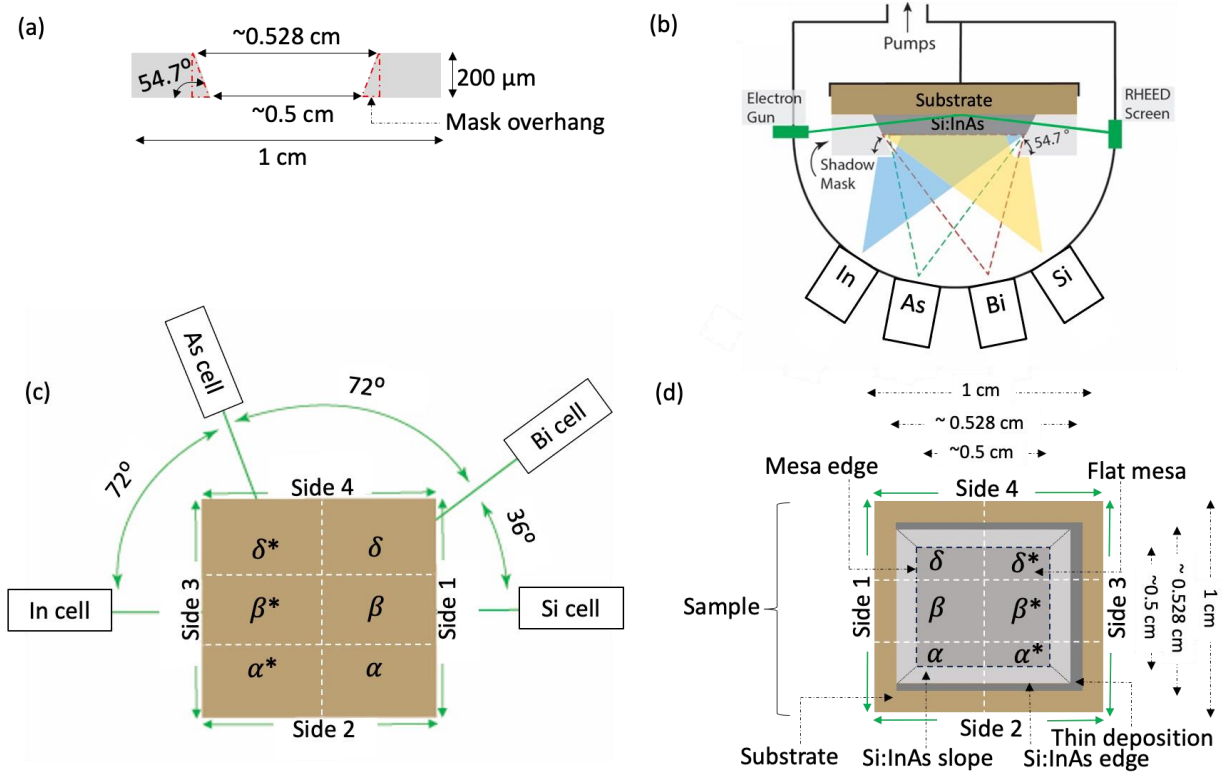

**Figure 1.** (a) Cross-sectional view of the silicon shadow mask in its upside-down orientation. (b) Sketch of MBE growth chamber showing cross-sectional view of the shadow mask, substrate, and Si:InAs film. (c) Top-down view of the positions of the cells with respect to the substrate during sample growth (showing back side of the substrate). (d) View of the Si:InAs film (showing front side of the substrate). The sample is divided into six regions: $\alpha$, $\beta$, $\delta$, $\alpha^*$, $\beta^*$ and $\delta^*$ for ease of discussion. Figures are not to scale.

The samples comprise a Si:InAs film grown on an unintentionally-doped GaSb (100) single side polished wafer from WaferTech using a Veeco GenXplor MBE system. Each wafer was diced into 1 cm × 1 cm substrates. During sample growth, the mask was placed on top of the substrate with the larger side of the mask opening in contact with the substrate, as shown in **Figure 1(b)**, and the substrate was not rotated. Side 1 of the substrate was fixed toward the silicon dopant cell which placed side 3 closest to the indium cell, enhancing the silicon and indium flux shadowing to obtain the largest in-plane permittivity gradient. The arsenic and bismuth cells were closer to side 4 than side 2, but they were not aligned with a substrate edge. A top-down view of the positions of the cells with respect to the substrate during growth is shown in **Figure 1(c)**, displaying the back side of the substrate (brown). **Figure 1(d)** shows the front side of the substrate (brown) and the Si:InAs film which is not in the center of the substrate but is offset toward side 1 due to the position of the indium cell. The Si:InAs film has a flat mesa (medium gray) with four mesa edges, as indicated in Figure 1(d). Below the mesa edge on each side, the thickness of the Si:InAs varies in the horizontal direction, giving rise to a Si:InAs slope (light gray). The bottom of the Si:InAs slope where the Si:InAs film touches the substrate is the Si:InAs edge, as indicated in Figure 1(d). Surrounding the Si:InAs edges, there is a layer of thin deposition on three sides of the sample (dark gray).

The sample is conceptually divided into six regions for ease of discussion, as shown in Figure 1(c) and 1(d). Regions $\alpha$, $\beta$ and $\delta$ are near side 1, closest to the silicon source. Regions $\alpha^*$, β* and δ* are near side 3, closest to the indium source. Regions α and α* are closest to side 2, while regions δ and δ* are closest to side 4, near the bismuth and arsenic sources. **Figure 2** shows optical micrographs of the sample taken from regions α, α*, δ, and δ*. During sample growth, side 4 was closest to the arsenic cell while side 2 was furthest from the arsenic cell; the arsenic beam was therefore shadowed for regions δ and δ*. This reduction in arsenic flux impacts the Si:InAs film quality near the mesa edge at side 4, resulting in a dark stripe as shown in Figure 2 (pink arrow).

In addition to the dark stripe, we see thin deposition surrounding the Si:InAs edges. The green dashed boxes in the figure indicate the areas where thin deposition occurs under the overhanging portion of the mask (orange arrow). The widths of these areas vary: side 3 is the widest (~300 μm), followed by side 4 and side 2. Side 1 has no noticeable deposition. This is because during growth, the indium cell was across from side 1, so there was no shadowing of the indium beam at side 1, resulting in no significant thin deposition and causing the mesa to be offset toward side 1. Side 3 was near the indium cell during growth, so the indium beam was almost fully shadowed for side 3, resulting in the widest area of thin deposition on side 3. The thin depositions occur under the mask overhang: the opening of the mask at the top is ~0.5 cm and at the bottom is ~0.528 cm, leading to mask overhangs which are each ~140 μm wide on each side of the mask (for details, see Figure 1(a) and Supporting Information). These mask overhangs give rise to the Si:InAs slopes as well as the thin deposition surrounding the Si:InAs edges. If we consider side 1 and side 3 together, the total mask overhang width is ~280 μm. This is comparable to the measured thin deposition width of ~300 μm on side 3. Similarly, thin depositions occur under the mask overhang on sides 2 and 4, which is caused by partial shadowing of the indium beam since the molecular beam from the cell is not perfectly collimated. The total width of the thin depositions on sides 2 and 4 together is ~260 μm, which is again comparable to the theoretical width of ~280 μm. The thin depositions on sides 2 and 4 are not of equal widths due to a slight misalignment of side 3 with respect to the indium cell.

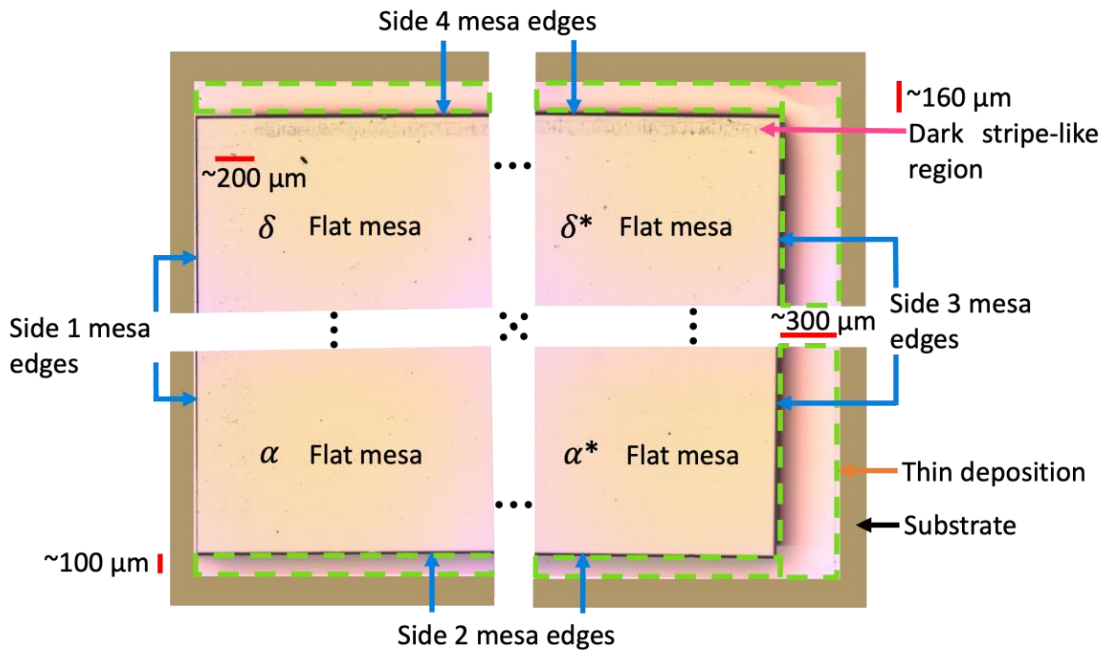

**Figure 2.** Optical micrographs of the sample from regions α, α*, δ and δ*. The bare substrate is colored brown in the image for clarity. The brown substrate area remains covered by mask during sample fabrication process. The blue

arrows indicate the mesa edges. The dark stripe along the mesa edge of side 4 is indicated by the pink arrow. Thin deposition surrounding the Si:InAs film are indicated by green dashed boxes and labelled by the orange arrow.

The relative positions of the elemental cells with respect to the substrate during sample growth impact the mesa edges, resulting in variation of the edge profiles, thicknesses, and compositions. **Figure 3(a)** shows surface profilometry scans of the sample in the regions $\alpha$, $\beta$, and $\delta$ on side 1 and $\alpha^*$, $\beta^*$, and $\delta^*$ on side 3. The approximate locations of the scans are indicated by the arrows in **Figure 3(b)**. We see that the mesa edge on side 1 (closest to silicon; farthest from indium) is not uniform along its length, leading to differences in surface quality. The profiles from regions $\beta$ and $\delta$ have large bumps indicative of rougher growth, in contrast to region $\alpha$ where we see a smoother surface. As described above, during growth, side 4 was closest to the arsenic cell while side 2 was furthest from the arsenic cell. This caused a reduction in arsenic flux for side 4, and an excess of arsenic flux for side 2. As a result, region $\alpha$ was exposed to an enhanced amount of arsenic in comparison to the neighboring regions $\beta$ and $\delta$, and as such, has a smoother surface. Compared to non-SMMBE growth of Si:InAs, the arsenic flux used for this sample was 1.5× higher to minimize the size of the arsenic-shadowed areas. Region $\beta$ was directly opposite to the indium source and had maximum exposure to the indium flux, which led to a lower As:In ratio and rougher growth in this region. Chemical analysis of region $\beta$ (**Figure 4(b)**) reveals that the bumps are primarily composed of indium. In contrast to side 1, the mesa edge at side 3 (closest to indium; furthest from silicon) is smoother along its length, as seen in the surface profiles from regions $\alpha^*$, $\beta^*$ and $\delta^*$. Unlike side 1, side 3 was shadowed from the indium beam during growth. This caused a reduction in the indium flux for side 3, increasing the As:In flux ratio and eliminating the indium droplets. However, region $\delta^*$ was shadowed from the arsenic flux, lowering the As:In flux ratio compared to the adjacent regions $\alpha^*$ and $\beta^*$ and resulting in slightly rougher growth.

We also observe a difference in the Si:InAs mesa slopes on sides 1 and 3: the slope is much steeper on side 1 in comparison to side 3. In particular, the slope on side 1 is ~466 ± 130 nm/μm and on side 3 is ~72 ± 5 nm/μm (see Supporting Information for the linear fits). This is due to the different shadowing of the indium cell. Side 1 was across from the indium cell and therefore had the least shadowing of the indium flux, leading to a steeper slope. Side 3 was closest to the indium cell and therefore experienced the largest shadowing of the indium flux, leading to a shallower slope.

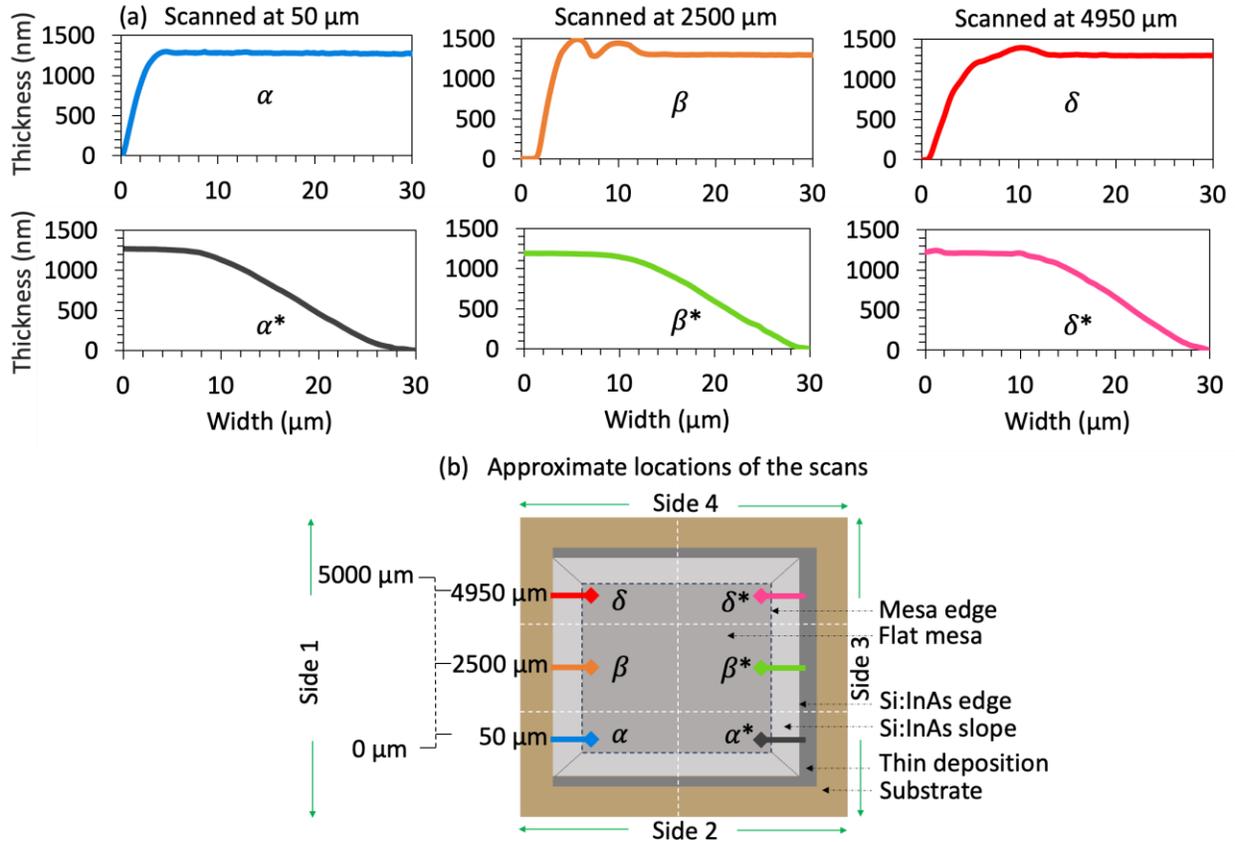

**Figure 3.** (a) Surface profiles of the sample from regions α, β, δ, α*, β* and δ*. (b) Approximate locations and directions of the surface profiles from Figure 3(a).

To improve our understanding of the surface morphologies of the sample, SEM images were taken from regions α, β and δ and are shown in **Figure 4(a)**. We observe trenches elongated along the [110] direction[36] in regions α and δ; these are marked with black dashed boxes. They appear near the mesa edge of side 2 in region α and near the mesa edge of side 4 in region δ. The trenches correspond to silicon surface segregation as described in references[36–38]. Near the mesa edge of side 1 in regions β and δ, sharp-looking structures are marked in white broken boxes. These structures are denser in region δ compared to region β, indicating that these features are likely related to the lower As:In flux ratio in this region. Finally, droplet features are observed at the mesa edge of side 1 in region β. Chemical analysis by EDS shown in Figure 4(b) reveals that these features are primarily composed of indium. Due to the positions of the indium and arsenic sources, region β had the lowest As:In flux ratio, likely resulting in indium droplets. Fortunately, our in-plane permittivity gradient regions are not near the mesa edge of side 1, but much further away; the GPM regions will be discussed later. **Figure 4(c)** shows SEM images of regions α*, β*, and δ*. We again observe trenches elongated along the [110] direction corresponding to silicon surface segregation[36–38]. Due to the positions of the silicon and indium cells, side 3 received a high silicon flux and a low indium flux, leading to excess silicon and promoting silicon surface segregation.

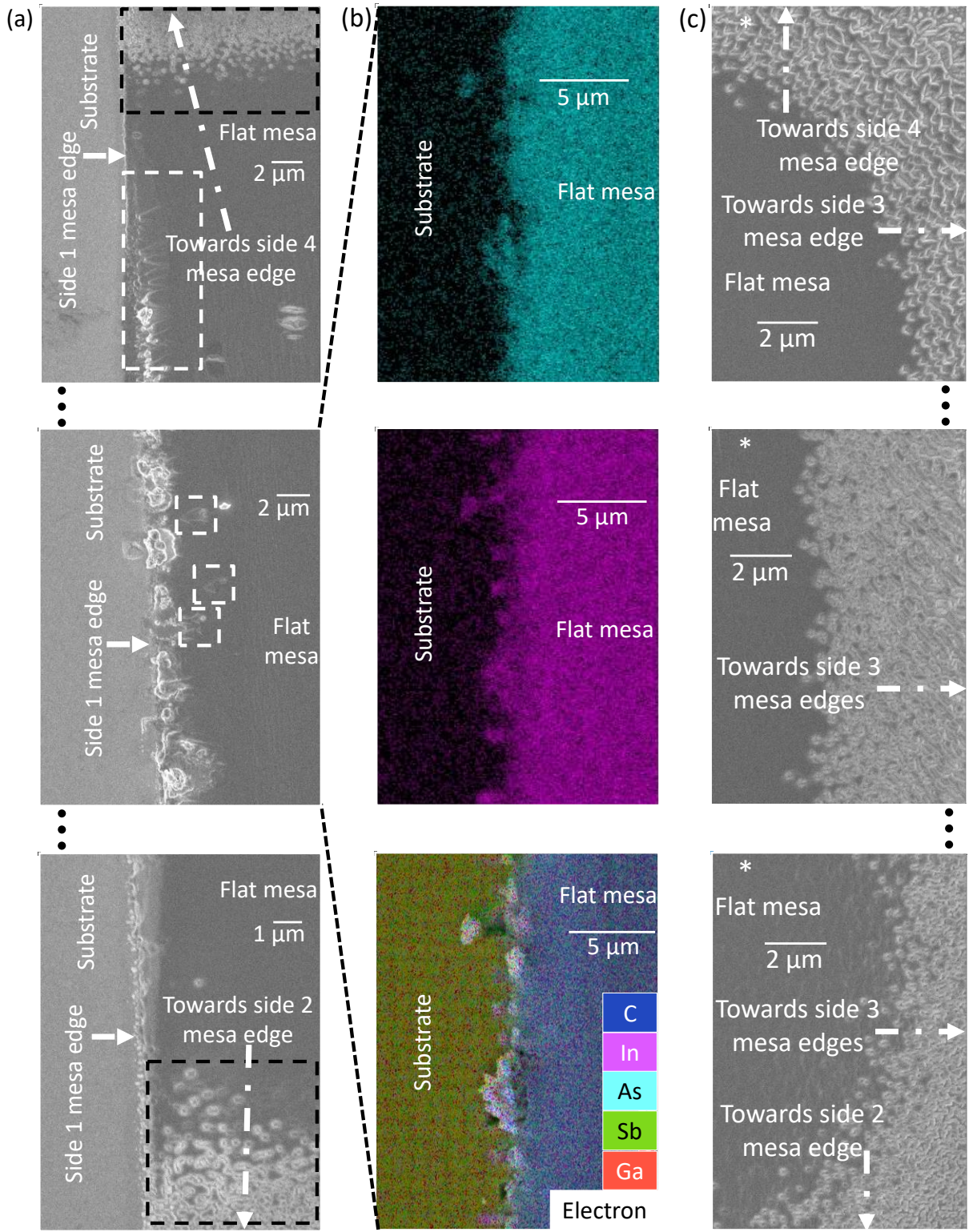

**Figure 4.** (a) Surface morphologies via SEM from regions $\alpha$, $\beta$ and $\delta$, respectively. (b) Chemical analysis via EDS spectroscopy from region $\beta$. (c) Surface morphologies via SEM from regions $\alpha^*$, $\beta^*$ and $\delta^*$, respectively. All the images in Figure 4(a-c) are rotated by 90° in the anti-clockwise direction to match the schematic of our sample orientation.

The goal of this project is to create an in-plane silicon doping gradient, giving rise to an in-plane permittivity gradient. We measured the silicon dopant concentration as a function of position using TOF-SIMS; these data are shown in the Supporting Information. To analyze the in-plane permittivity, we used nano-FTIR. **Figure 5** shows nano-FTIR spectra as a function of position from regions $\alpha$, $\beta$ and $\delta$ as well as the corresponding surface roughness. Each scan was initiated ~500 μm away from the side 1 mesa edge on the flat portion of the mesa; details are shown in the supporting information. The scans are toward the center of the mesa. To avoid the dark stripe shown in the optical micrographs in Figure 2, the scan from region $\delta$ was taken ~150 μm away from the side 4 mesa edge. For consistency, the scan from region $\alpha$ was also taken ~150 μm away from the side 2 mesa edge. In all spectra (Figure 5), we see localization and enhancement of different light wavelengths at different positions. Specifically, we see electric field enhancements corresponding to a *wavenumber gradient* of ~650 cm$^{-1}$ to 900 cm$^{-1}$ over an *in-plane gradient width* of ~13 μm, indicating the successful creation of an in-plane GPM.

To quantify the permittivity gradients, we define two parameters: the *wavenumber gradient width* is the difference between the maximum and minimum wavenumbers, and the *wavenumber gradient steepness* is the ratio of wavenumber gradient width to the in-plane gradient width. For all three regions, the wavenumber gradient width is ~250 cm$^{-1}$ and the wavenumber gradient steepness is ~19.2 cm$^{-1}$/μm. These spectra from regions $\alpha$, $\beta$ and $\delta$ were all obtained on the flat mesa, indicating that the permittivity gradient comes from the variation in silicon flux rather than the variation in indium flux. This matches our expectations, given that this side of the sample had minimal shadowing of the indium flux and maximal shadowing of the silicon flux. From the surface roughness values, we see that region $\delta$ has the highest surface roughness ($R_{rms}$ ~1.1 nm), followed by region $\beta$, followed by region $\alpha$. As mentioned previously, due to the positions of the cells, region $\alpha$ had the largest As:In flux ratio, followed by $\beta$ and $\delta$, leading directly to differences in surface roughness.

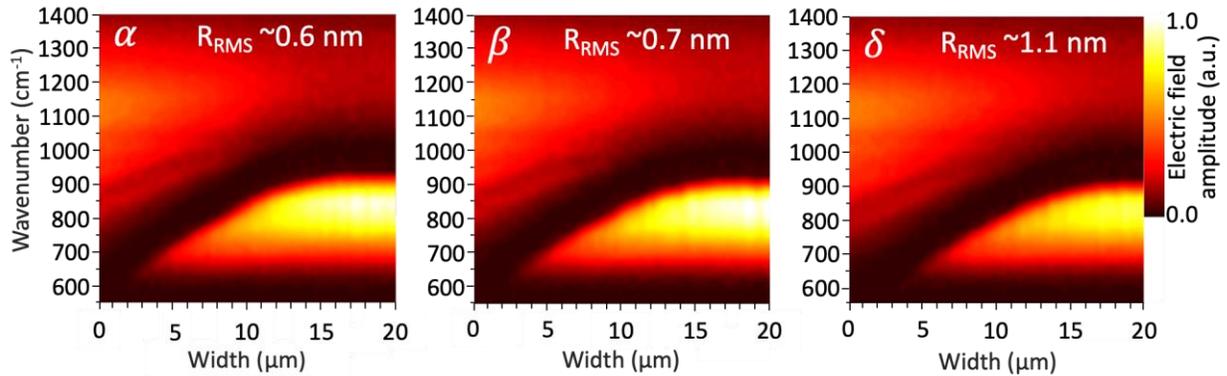

**Figure 5.** Nano-FTIR spectra from regions $\alpha$, $\beta$ and $\delta$, and the corresponding RMS surface roughness via s-SNOM. The approximate scan locations are shown in the Supporting Information.

Unlike the gradients in regions $\alpha$, $\beta$, and $\delta$ on side 1, the gradients in regions $\alpha^*$, $\beta^*$, and $\delta^*$ are not in the flat mesa areas; instead, they are on the Si:InAs slope on side 3. Due to the positions of the silicon and indium cells, side 3 received a high silicon flux and a low indium flux. Therefore, the permittivity gradient in this case arises primarily from the variation in indium flux rather than the variation in silicon flux: the silicon flux as a function of position is constant while the indium flux decreases as a function of position in the Si:InAs slope below the mesa edge. This leads to a concomitant increase in the silicon doping density and gives rise to gradient in permittivity. **Figure 6** shows the nano-FTIR spectra as a function of position from regions $\alpha^*$, $\beta^*$ and $\delta^*$, and the corresponding surface profiles of the scan locations with their surface roughness. Each of these scans started on the flat mesa near the mesa edge, moved down the Si:InAs slope, and ended on the thin deposition at the bottom of the Si:InAs edge (see Supporting Information for the approximate locations of the scans). The scan from region $\alpha^*$ was obtained ~150 μm away from the mesa edge at side 2, like that of region $\alpha$; the scan from region $\delta^*$ was also obtained ~150 μm away from the mesa edge at side 4, like that of region $\delta$. None of the spectra show a wavenumber gradient in the flat mesa area. However, an electric field enhancement corresponding to a wavenumber gradient is observed in each of the spectra as the sample thickness varies down the Si:InAs slope. Region $\beta^*$ has a wavenumber gradient of ~900 cm$^{-1}$ to 1250 cm$^{-1}$ over an in-plane gradient width of ~13 μm, giving a wavenumber gradient width of ~350 cm$^{-1}$ and a wavenumber gradient steepness of ~26.9 cm$^{-1}$/μm. Region $\alpha^*$ has a wavenumber gradient width of ~250 cm$^{-1}$ and a wavenumber gradient steepness of ~35.7 cm$^{-1}$/μm, while region $\delta^*$ has a wavenumber gradient width of ~250 cm$^{-1}$ and a wavenumber gradient steepness of ~41.7 cm$^{-1}$/μm. In contrast to regions $\alpha$, $\beta$, and $\delta$, the permittivity gradients in regions $\alpha^*$, $\beta^*$, and $\delta^*$ are steeper and wider and reach a higher maximum wavenumber of ~1250 cm$^{-1}$. However, after reaching a maximum wavenumber of ~1250 cm$^{-1}$, the spectra become extremely noisy. This is likely because the silicon doping density has increased beyond the maximum doping density of InAs, resulting in over-doped areas.

The surface profiles of the scanned locations in Figure 6 display the surface roughness on the mesa area and the thin deposition area separately. Like the regions on side 1, the surface roughness of regions $\alpha^*$, $\beta^*$, and $\delta^*$ on side 3 is influenced by the As:In ratio. Region $\delta^*$ had the lowest As:In ratio and has the highest roughness on the mesa, followed by regions $\beta^*$ and $\alpha^*$. The thin deposition on side 3 is an area that was under the mask overhang during sample growth, hence this area is very rough in comparison to the mesa regions. The surface roughness of the thin depositions in regions $\alpha^*$, $\beta^*$, and $\delta^*$ are similar at ~30 nm.

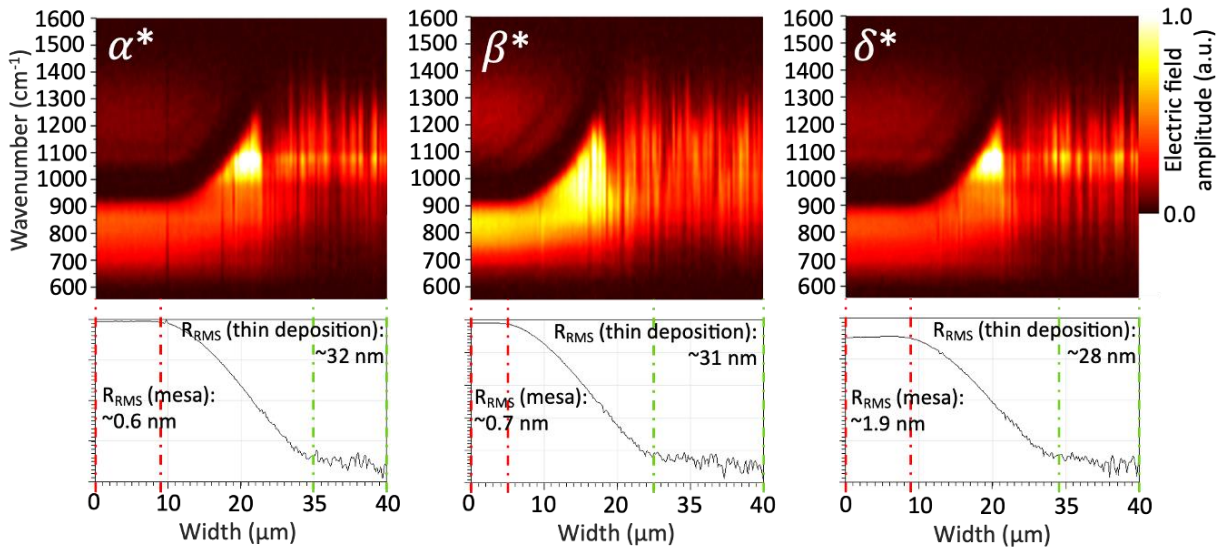

**Figure 6.** Nano-FTIR spectra from regions $\alpha^*$, $\beta^*$ and $\delta^*$, and the corresponding surface profiles via s-SNOM (see approximate scan locations in the Supporting Information). The surface roughness of the mesa area and the thin deposition area were measured separately. Here, the mesa area is denoted by the red broken lines, while the thin deposition area is denoted by the green broken lines. The RMS surface roughness of the areas is denoted by $R_{RMS}$.

## 3. Conclusions

In conclusion, we have used SMMBE to make in-plane permittivity gradients in Si:InAs thin films. Nano-FTIR spectra as a function of position from our SMMBE Si:InAs film show localization and enhancement of different light wavelengths at different horizontal locations on the film, confirming the formation of in-plane permittivity gradients along the film width. We obtain in-plane permittivity gradients on two opposite sides of the film – on the flat mesa on one side with silicon shadowing and on the film slope on the opposite side with indium shadowing. The material quality of the film depends on As:In flux ratio. Regions which had a lower As:In flux ratio showed rougher growth, while regions which had a higher As:In flux ratio showed smoother growth.

Both the in-plane permittivity gradients on the flat mesa and the slope can be used to make devices, such as an on-chip ultracompact spectrometer. However, it is more convenient to work on a flat area in comparison to an area on a slope for device fabrication purposes. In future, the in-plane permittivity gradient width and steepness can be tailored by tailoring the mask design parameters, enabling the design of GPMs tailored to the specific device application.

## 4. Experimental methods

To synthesize our samples, we used removable and potentially reusable shadow masks made of silicon. The first SMMBE trial using as-received mask without cleaning resulted in significant contamination of the sample, leading us to institute a cleaning process immediately prior to use. Prior to loading in the MBE chamber, the masks were cleaned with solvents. We used two different cleaning methods *method 1* and *method 2*, as described in **Table 1**. Method 1 worked well, leading to clean masks and no obvious sample contamination. However, over multiple trials, a few masks broke during the cleaning process, likely due to the sonication. As a result, we tried method 2, which uses the same solvents in the same order, but without sonication. Method 2 also worked well, leading to clean masks and no obvious sample contamination; no masks broke using method 2.

**Table 1.** Step-by-step description of two different mask cleaning methods.

| Method 1 | Method 2 |
| --- | --- |
| Sonicate in acetone at room temperature for 1 minute | Spray each side of the mask with acetone |
| Sonicate in isopropanol alcohol (IPA) at room temperature for 1 minute | Spray each side of the mask with isopropanol alcohol (IPA) |
| Blow dry each side of the mask with nitrogen gun | Blow dry each side of the mask with nitrogen gun |
| Bake each side of the mask on a hot plate at 280 °C | Bake each side of the mask on a hot plate at 280 °C |

Prior to film deposition, the substrate was thermally deoxidized under an antimony overpressure. The Si:InAs film was then directly nucleated on top of the substrate at a substrate temperature of 500 °C. We used a bismuth surfactant during the Si:InAs deposition to improve the incorporation of silicon into InAs and to reduce the optical scattering rate.[36] The substrate was rotated during deoxidation, but rotation was turned off during Si:InAs deposition to fix side 1 toward the silicon dopant cell. The silicon dopant cell was a filament cell from MBE Komponenten which runs current through a silicon filament to heat the filament and evaporate silicon. This silicon flux was controlled by the filament current, and the temperature of the filament was measured by a non-contact thermocouple. We have found that constant current mode sometimes shows temperature oscillations, and the temperature of the cell drops by ~11 °C over ~3 hours. The silicon cell current was set at 44 A, and the cell temperature fluctuated between 1032 °C and 1046 °C. The growth rate was ~1.89 μm/hr, and the growth time was ~45 minutes. The beam equivalent pressure (BEP) of bismuth and arsenic were ~5% indium BEP and ~30× indium BEP, respectively.

After growth, the sample was characterized to determine its physical and optical properties. Optical micrographs were captured with a Zeiss Axio Imager Z2 Vario Microscope. Profilometry scans were used to determine the thickness gradient with a Bruker DektakXT Stylus Profiler. Line scan to measure the intensity of the silicon dopant across the sample was taken with an IONTOF TOF.SIMS[5] time-of-flight secondary ion mass spectrometry (TOF-SIMS). A Neaspec s-SNOM (scattering-type scanning nearfield optical microscope) with a mid-IR module (~650 $cm^{-1}$ – 1400 $cm^{-1}$) was used to acquire nanoscale Fourier transform infrared (Nano-FTIR) spectra as a function of position to measure the in-plane permittivity gradient using $4^{th}$ order optical amplitudes along a line on the sample. The Neaspec s-SNOM is also equipped with an AFM unit which was used to acquire 5 μm × 5 μm area scans for surface roughness values from the approximate locations on the film from where the nano-FTIR spectra were attained. The same AFM unit was used to get the surface profile trends from the approximate locations as the nano-FTIR spectra. Scanning electron microscope (SEM) images were taken to understand the surface morphology of the sample using a Zeiss MERLIN SEM. Energy-dispersive X-ray spectroscopy (EDS) was done to perform chemical analysis on the sample using a Zeiss Auriga 60 SEM tool.

**Supporting Information**

Supporting Information is available from the Wiley Online Library or from the author.

**Acknowledgements**

This material is based upon work supported by the U.S. National Science Foundation under award no. ECCS-2102027, the use of facilities and instrumentation supported by the U.S. National Science Foundation through the University of Delaware Materials Research Science and Engineering Center under award no. DMR-2011824, and the use of TOF-SIMS sponsored by the U.S. National Science Foundation, Major Research Instrumentation under award no. DMR-2116754.

**References**

[1] T. Schallenberg, *Shadow Mask assisted Heteroepitaxy of Compound Semiconductor Nanostructures*, **2004**.
[2] G. Kaminsky, *Journal of Vacuum Science & Technology B: Microelectronics and Nanometer Structures* **1985**, *3*, 741.
[3] Y. Luo, L. Zeng, W. Lin, B. Yang, M. C. Tamargo, Y. M. Strzhemechny, S. A. Schwarz, *J Electron Mater* **2000**, *29*.
[4] A. Y. Cho, F. K. Reinhart, *Appl Phys Lett* **1972**, *21*, 355.

[5]     W. T. Tsang, M. Ilegems, *Appl Phys Lett* **1977**, *31*, 301.
[6]     Y. Luo, *Journal of Vacuum Science & Technology B: Microelectronics and Nanometer Structures* **1998**, *16*, 1312.
[7]     A. Y. Cho, P. D. Dernier, *J Appl Phys* **1978**, *49*, 3328.
[8]     C. Schumacher, W. Faschinger, V. Hock, H. R. Reß, J. Nürnberger, M. Ehinger, *J Cryst Growth* **1999**, *201*.
[9]     K. H. Gulden, X. Wu, J. S. Smith, P. Kiesel, A. Höfler, M. Kneissl, P. Riel, G. H. Döhler, *Appl Phys Lett* **1993**, *62*, 3180.
[10]    T. Schallenberg, C. Schumacher, W. Faschinger, *In situ structuring during MBE regrowth with shadow masks*, Vol. 13, **2002**.
[11]    P. B. Welander, V. Bolkhovsky, T. J. Weir, M. A. Gouker, W. D. Oliver, *Shadow evaporation of epitaxial Al/Al2O3/Al tunnel junctions on sapphire utilizing an inorganic bilayer mask*, **2012**.
[12]    T. Schallenberg, L. W. Molenkamp, S. Rodt, R. Seguin, D. Bimberg, G. Karczewski, *Appl Phys Lett* **2004**, *84*, 963.
[13]    T. Schallenberg, W. Faschinger, G. Karczewski, L. W. Molenkamp, V. Türck, S. Rodt, R. Heitz, D. Bimberg, M. Obert, G. Bacher, A. Forchel, *Appl Phys Lett* **2003**, *83*, 446.
[14]    T. Schallenberg, C. Schumacher, K. Brunner, L. W. Molenkamp, In *Physica Status Solidi (B) Basic Research*, **2004**, pp. 564–567.
[15]    T. Schallenberg, C. Schumacher, L. W. Molenkamp, In *Physica Status Solidi (A) Applied Research*, **2003**, pp. 232–237.
[16]    T. Schallenberg, T. Borzenko, G. Schmidt, L. W. Molenkamp, S. Rodt, R. Heitz, D. Bimberg, G. Karczewski, *J Appl Phys* **2004**, *95*, 311.
[17]    T. Schallenberg, T. Borzenko, G. Schmidt, M. Obert, G. Bacher, C. Schumacher, G. Karczewski, L. W. Molenkamp, S. Rodt, R. Heitz, D. Bimberg, *Appl Phys Lett* **2003**, *82*, 4349.
[18]    T. Schallenberg, K. Brunner, T. Borzenko, L. W. Molenkamp, G. Karczewski, *Surface diffusion during shadow-mask-assisted molecular-beam epitaxy of III-V compounds*, Vol. 98, **2005**.
[19]    T. Schallenberg, L. W. Molenkamp, G. Karczewski, *Phys Rev B Condens Matter Mater Phys* **2004**, *70*.
[20]    N. Tomita, N. Yoshida, S. Shimomura, K. Murase, A. Adachi, S. Hiyamizu, *J Cryst Growth* **1995**, *150*, 377.
[21]    H. Gao, E. Herrmann, X. Wang, *Opt Express* **2020**, *28*.
[22]    S. Law, D. C. Adams, A. M. Taylor, D. Wasserman, *2012 IEEE Photonics Conference, IPC 2012* **2012**, *20*, 786.
[23]    D. Wei, C. Harris, C. C. Bomberger, J. Zhang, J. Zide, S. Law, *Opt Express* **2016**, *24*.
[24]    J. R. Felts, S. Law, C. M. Roberts, V. Podolskiy, D. M. Wasserman, W. P. King, *Appl Phys Lett* **2013**, *102*.
[25]    P. Sohr, D. Wei, S. Tomasulo, M. K. Yakes, S. Law, *ACS Photonics* **2018**, *5*.
[26]    P. Sohr, C. I. Ip, S. Law, *Opt Lett* **2019**, *44*.
[27]    S. Law, L. Yu, D. Wasserman, *Journal of Vacuum Science & Technology B, Nanotechnology and Microelectronics: Materials, Processing, Measurement, and Phenomena* **2013**, *31*, 03C121.
[28]    S. Law, R. Liu, D. Wasserman, *Journal of Vacuum Science & Technology B, Nanotechnology and Microelectronics: Materials, Processing, Measurement, and Phenomena* **2014**, *32*, 052601.

[29]     D. Wei, C. Harris, S. Law, *Opt Mater Express* **2017**, *7*.
[30]     A. J. Cleri, M. He, J. D. Caldwell, J. P. Maria, In *International Conference on Metamaterials, Photonic Crystals and Plasmonics*, **2022**.
[31]     H. Dittrich, A. Stadler, D. Topa, H. J. Schimper, A. Basch, In *Physica Status Solidi (C) Current Topics in Solid State Physics*, **2009**.
[32]     A. Rosenberg, J. Surya, R. Liu, W. Streyer, S. Law, L. Suzanne Leslie, R. Bhargava, D. Wasserman, *Journal of Optics (United Kingdom)* **2014**, *16*.
[33]     X. Ou, K. H. Heinig, R. Hübner, J. Grenzer, X. Wang, M. Helm, J. Fassbender, S. Facsko, *Nanoscale* **2015**, *7*.
[34]     J. Chan, T. Fu, N. W. Cheung, J. Ross, N. Newman, M. Rubin, *Comparison of AIN Films Grown by RF at Magnetron Sputtering and Ion-Assisted Molecular Beam Epitaxy*, **1994**.
[35]     Norcada.
[36]     D. Wei, S. Maddox, P. Sohr, S. Bank, S. Law, *Opt Mater Express* **2020**, *10*.
[37]     S. Muto, S. Takeda, M. Hirata, K. Fujii, K. Ibe, *Philosophical Magazine A: Physics of Condensed Matter, Structure, Defects and Mechanical Properties* **1992**, *66*.
[38]     C. Domke, P. Ebert, M. Heinrich, K. Urban, *Phys Rev B Condens Matter Mater Phys* **1996**, *54*.

# Shadow Mask Molecular Beam Epitaxy for In-Plane Gradient Permittivity Materials

S. Mukherjee[1], S. R. Sitaram[1], X. Wang[1], and S. Law[2,3,4,5,*]

[1]Department of Materials Science and Engineering, University of Delaware, 201 Dupont Hall, 127 The Green, Newark, DE 19716, USA

[2]Department of Materials Science and Engineering, The Pennsylvania State University, University Park, PA 16802, USA

[3]Materials Research Institute, The Pennsylvania State University, University Park, PA 16802, USA

[4]2D Crystal Consortium Materials Innovation Platform, The Pennsylvania State University, University Park, PA 16802, USA

[5]Penn State Institute of Energy and the Environment, The Pennsylvania State University, University Park, PA 16802, USA

[*]sal6149@psu.edu

## I. The combined geometry of mask and sample

**Figure S1** shows the combined geometry of the mask and the sample, with the mask resting on the front of the substrate. The figure displays a cross-sectional view of the mask and the sample from side 1 to side 3. The mask thickness is 200 μm, the mask size is 1 cm × 1 cm, the aperture opening at the top is 0.5 cm × 0.5 cm, and the angle of the aperture sidewalls is 54.7°. Using this information, the dimension of the aperture opening at the bottom of the mask and the width of the mask overhangs are calculated, as shown in Figure S1. From the cross-sectional schematic, we can see that the Si:InAs film is offset toward side 1 due to the position of the indium cell during growth. This results in the thin deposition only under the mask overhang at side 3 and not under the mask overhang at side 1. As discussed in the main text, the thin deposition also occurs under the mask overhangs at both sides 2 and 4 of the sample.

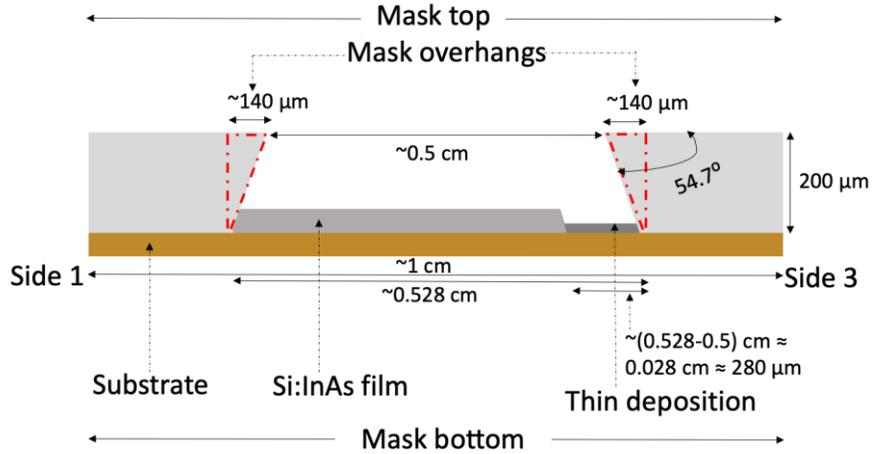

**Figure S1.** Combined geometry of the mask and the sample with the mask resting on the front of the substrate. The figure displays a cross-sectional view of the mask and the sample from side 1 to side 3. Figure is not to scale.

## II. Linear fit of Si:InAs film surface profiles

**Figure S2** shows the linear fits from the linear portions of the surface profiles of the mesa edges (**Figure 3(a)** in the main text) from regions $\alpha$, $\beta$, and $\delta$ on side 1, and $\alpha^*$, $\beta^*$, and $\delta^*$ on side 3. The slopes on side 1 are steeper and vary more in comparison to the slopes on side 3. The steepness of the slope on side 1 on average is ~466 ± 130 nm/μm and on side 3 on average is ~72 ± 5 nm/μm. The error bars in the average values of sides 1 and 3 is the standard deviation.

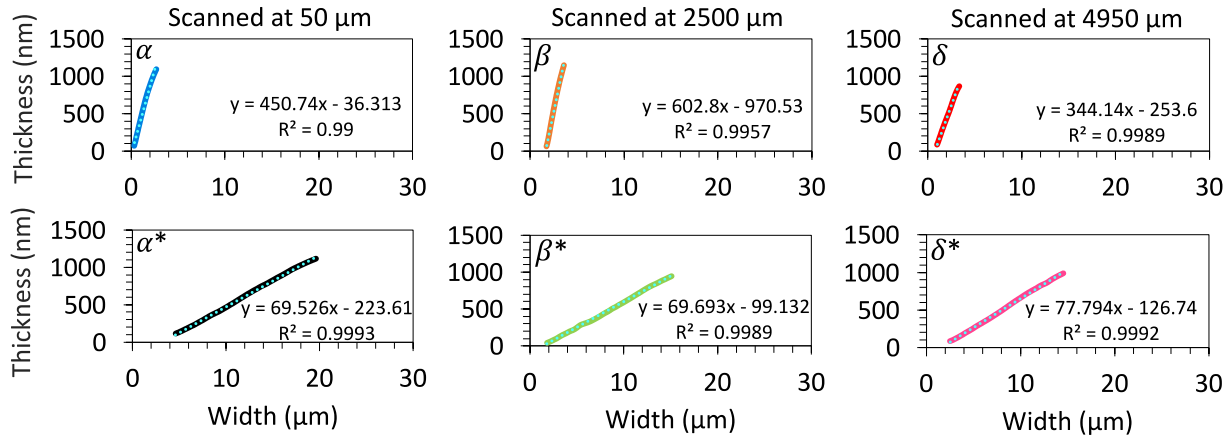

**Figure S2.** Linear fits from the linear portions of the surface profiles from regions $\alpha$, $\beta$, and $\delta$ at side 1, and $\alpha^*$, $\beta^*$, and $\delta^*$ at side 3. The solid lines are the experimental data; the dotted lines are the fits.

## III. Secondary ion mass spectroscopy

The aim of our work was to create a silicon doping gradient in the horizontal direction, giving rise to in-plane gradient in permittivity. We used TOF-SIMS line scans to measure the silicon dopant intensity as a function of position. **Figure S3(a)** shows an approximate location of the line scan on the sample as denoted by the location of the broken yellow arrow. The direction of the arrow denotes the scan direction. The scan was initiated at side 1 and was terminated at side 3, covering regions $\beta$ and $\beta^*$ of the sample. At both sides, the scan ran over substrate areas which were in contact with the mask during sample growth. **Figure S3(b)** displays the silicon intensity as a function of position along in-plane sample width. From the overall trend of the silicon intensity, we see that the intensity is very low in region $\beta$, and increases to a maximum value in region $\beta^*$. **Figure S3(c)** shows a zoomed in view (orange box) of the scan in Figure S3(b) from region $\beta$ near side 1. Here, we see the silicon intensity increases gradually from ~260 counts to ~540 counts. Side 1 was closest to the silicon dopant cell during sample growth, hence silicon beam was shadowed for side 1 near the mesa edge, causing a reduction in silicon flux for side 1. As a result, the silicon flux gradually increased toward the inward direction of the flat mesa from side 1. **Figure S3(d)** shows the zoomed in view (green box) of the scan in Figure S3(b) from region $\beta^*$ near side 3. Since the silicon cell was across from side 3 during sample growth, we see very high intensity of silicon in region $\beta^*$. Here the silicon intensity reaches a maximum value of ~32000 counts and then drops to ~1000 counts.

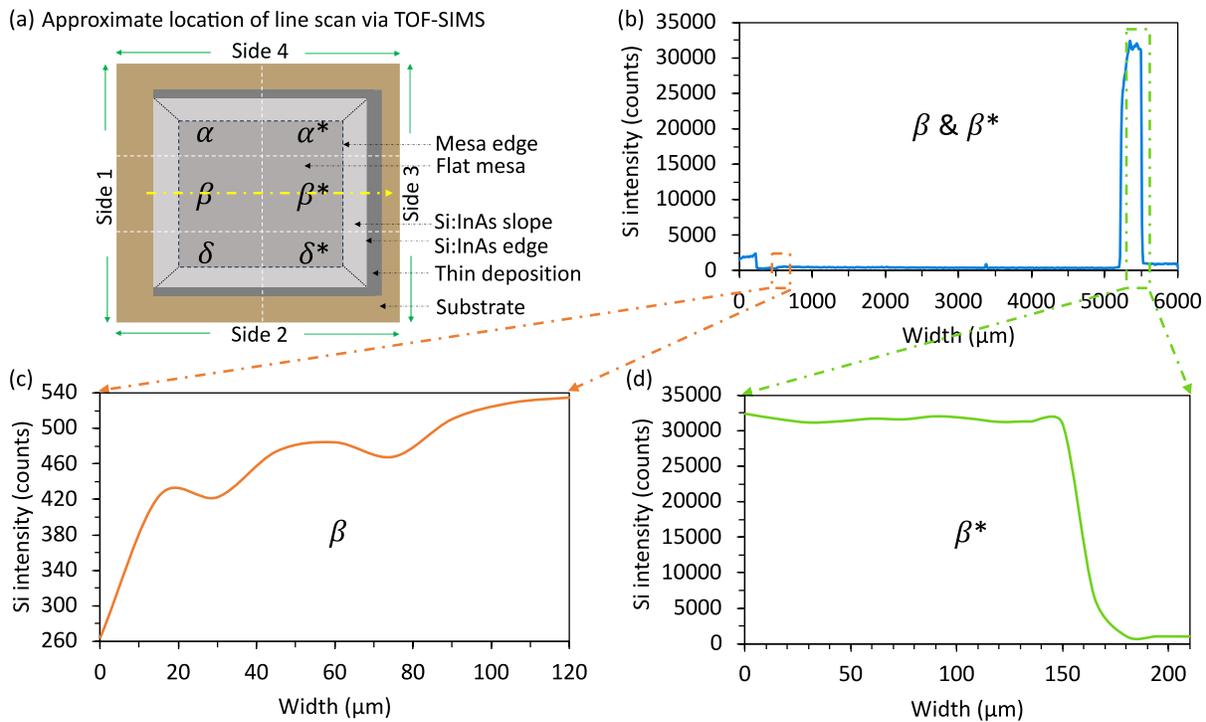

**Figure S3.** (a) Approximate location and direction of the TOF-SIMS line scan shown in Figure 3(b) (as indicated by the location and direction of the yellow broken arrow). (b) Line scan via TOF-SIMS from regions $\beta$ and $\beta^*$, representing silicon intensity as a function of position along in-plane sample width. (c) Silicon intensity as a function of position along in-plane sample width from region $\beta$. It is a zoomed-in portion from the scan in Figure 3(b) (marked by the orange dashed box). (d) Silicon intensity as a function of position along in-plane sample width from region $\beta^*$. It is a zoomed-in portion from the scan in Figure 3(b) (marked by the green dashed box).

## IV. Locations of s-SNOM scans for the nano-FTIR spectra

The main text shows nano-FTIR spectra from regions $\alpha$, $\beta$ and $\delta$ in **Figure 5**, and regions $\alpha^*$, $\beta^*$ and $\delta^*$ in **Figure 6**. **Figure S4** shows the approximate locations and directions of the scans in Figure. 5 and 6 in the paper. In regions $\alpha$, $\beta$ and $\delta$, the scans were initiated ~500 μm away from the mesa edge on side 1. In regions $\alpha^*$, $\beta^*$ and $\delta^*$, the scans started on the flat mesa near the mesa edge on side 3, moved down the Si:InAs slope, and ended on the thin deposition at the bottom of the Si:InAs edge.

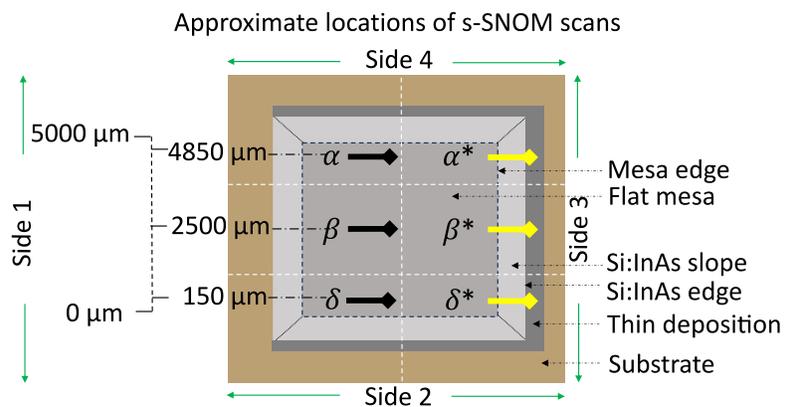

**Figure S4.** Approximate locations and directions of nano-FTIR scans in Figure. 5 and 6 of the main text.

Table of Content

# Shadow Mask Molecular Beam Epitaxy for In-Plane Gradient Permittivity Materials

S. Mukherjee[1], S. R. Sitaram[1], X. Wang[1], and S. Law[2,3,4,5,*]

[1]Department of Materials Science and Engineering, University of Delaware, 201 Dupont Hall, 127 The Green, Newark, DE 19716, USA

[2]Department of Materials Science and Engineering, The Pennsylvania State University, University Park, PA 16802, USA

[3]Materials Research Institute, The Pennsylvania State University, University Park, PA 16802, USA

[4]2D Crystal Consortium Materials Innovation Platform, The Pennsylvania State University, University Park, PA 16802, USA

[5]Penn State Institute of Energy and the Environment, The Pennsylvania State University, University Park, PA 16802, USA

[*]sal6149@psu.edu

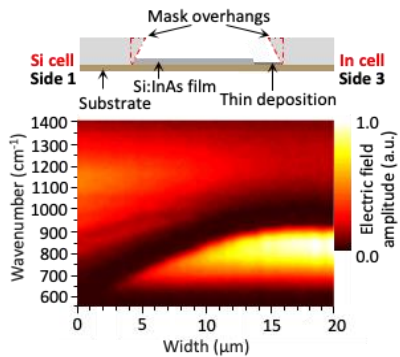